\documentclass[prb,twocolumn,floats,aps]{revtex4}
\newcommand{\kk}{{\bf k}}
\newcommand{\rr}{{\bf r}}
\usepackage{graphicx}
\usepackage{epstopdf}
\begin{document}

\title{The geometrically-averaged density of states as
a measure of localization}
\author{Yun Song, W. A. Atkinson, R. Wortis}
\affiliation{Department of Physics \& Astronomy, Trent University,
1600 West Bank Dr., Peterborough ON, K9J 7B8, Canada}
\date{\today}

\begin{abstract}
Motivated by current interest in disordered systems of interacting
electrons, the effectiveness of the geometrically averaged density of
states, $\rho_g(\omega)$, as an order parameter for the Anderson
transition is examined.  In the context of finite-size systems we
examine complications which arise from finite energy resolution.
Furthermore we demonstrate that even in infinite systems a decline in
$\rho_g(\omega)$ with increasing disorder strength is not uniquely
associated with localization.
\end{abstract}
\maketitle

\section{Introduction}
\label{sec-intro}

Although disordered metallic systems have been studied extensively for
many years, a reasonably complete theoretical understanding 
has only been developed for noninteracting particles,\cite{lee:85}
and the effect of interactions in disordered media is an open and
challenging problem.  

In 1958, Anderson\cite{anderson:58} proposed that noninteracting electrons may be localized by 
the interference of wavefunctions elastically scattered off of multiple impurities
even in situations in which the electrons
are classically unconfined.  Extensive studies\cite{lee:85} since then
have produced a clear picture of the Anderson transition.  In $d=3$
dimensions, and for sufficient disorder, single-particle wavefunctions
at the band edges are localized, and decay exponentially on a scale
referred to as the localization length, $\xi$.  $\xi$ is a function of
energy, and diverges at the upper and lower mobility edges
$\omega_{c\pm}$ which separate localized and extended states; states
with energies $\omega > \omega_{c+}$ or $\omega < \omega_{c-}$ are
localized, while states with $\omega_{c-} < \omega < \omega_{c+}$ are
extended.  The mobility edges move towards the band center as disorder
is increased, and a metal-insulator transition occurs when either
mobility edge crosses the Fermi level.

The question of how electron-electron interactions influence the
interference effects responsible for Anderson localization has
received much attention \cite{lee:85,belitz:94,abrahams:01} with
calculations using mean field, diagrammatic perturbation and
renormalization techniques.  While much has been learned, a clear
understanding of the strong interaction limit has not been achieved.
Recently a number of approaches have been developed to use dynamical
mean field theory (DMFT) methods to study the Mott-Anderson
metal-insulator transiton.  \cite{ dobrosavljevic:94,
dobrosavljevic:97, aguiar:03, dobrosavljevic:03, byczuk:05,
miranda:05, aguiar:06} In its original form DMFT is a local
approximation in which a single site is embedded in an effective
medium.  The distinction from standard mean-field theory is that the
on-site interaction is treated exactly, and the effective medium is
chosen to approximate the effects of other interactions.  One solves
self-consistently for the local single-particle Green's function
$G(\rr,\rr,z)$, where $\rr$ is a position and $z$ a complex frequency.

DMFT captures the physics of the interaction-driven metal-insulator
transition\cite{georges:96} (the Mott transition), but two significant
challenges arise in its application to the study of the Anderson
transition.  First, one must capture the subtle physics of
localization
within an effective medium theory and, second, one must extract
information about nonlocal behavior from the local Green's function.
The first problem has been the topic of recent work, \cite{
dobrosavljevic:94, dobrosavljevic:97, aguiar:03, dobrosavljevic:03,
byczuk:05, miranda:05, aguiar:06} while the focus of the current
article is the second: What quantity calculable from the local Green's
function is a practical measure of localization?

For numerical calculations on noninteracting-electron systems, a
number of convenient measures exist: including the Thouless
number,\cite{edwards:72,thouless:74} spectral
statistics,\cite{hofstetter:93} and especially the inverse
participation ratio\cite{bell:70,thouless:74} (IPR).  A key point,
however, is that in the absence of electron-electron interactions the
many-body wavefunction is a product of well-defined single particle
eigenstates, and all of these methods rely on knowledge of these
states.  However, in interacting systems no such states exist.

A possible approach is suggested by the
fact\cite{thouless:70,thouless:74} that for a system with localized
states the local density of states (LDOS), $\rho(\rr,\omega) =
-\pi^{-1} \mbox{Im }G(\rr,\rr,\omega+i0^+)$, has a discrete spectrum.
In an infinite system the density of states of the whole system (DOS)
is in general a continuous function of $\omega$, and when the states
are extended the LDOS is correspondingly continuous.  However, in an
infinite system with localized states the DOS will remain continuous
while the LDOS becomes discrete.  \cite{thouless:70,abouchacra:73,thouless:74} 
This has been discussed more recently in the context of non-interacting electrons by a
number of authors.\cite{mirlin:91,mirlin:94,janssen:98,brndiar:06}

With this in mind, 
the geometric average of the density of states (GADOS)
\begin{equation}
\rho_g(\omega) = \left [ \prod_{i=1}^N \rho(\rr_i,\omega) \right ]^{1/N},
\label{gados}
\end{equation}
has been proposed
\cite{
dobrosavljevic:97,
janssen:98,
aguiar:03, dobrosavljevic:03, byczuk:05} as a possible order parameter
of the Anderson localization transition.  Note that if the LDOS at a
given energy $\omega$ is zero on {\it any} site, the GADOS at that
energy will be zero.  Therefore, in an infinite system $\rho_g(0)
\rightarrow 0$ signals a metal-insulator transition, where $\omega=0$
corresponds to the Fermi level.

In this paper we explore some potential pitfalls in the use of this
measure.  Our central results are (1) that a finite energy resolution 
changes the scaling behavior of $\rho_g(\omega)$
and makes the determination of the mobility edges difficult
and (2) that simply a decline, as opposed to a zero, in the GADOS is not uniquely
associated with the phenomenon of localization.  We arrive at these
conclusions in the context of a system of non-interacting electrons on
a lattice with random potential values at each site (Section
\ref{ssec-model}).  First, we consider finite-size systems and use a
standard numerical diagonalization approach (Section \ref{ssec-enc}).
In these systems, the density of states is discrete, but a spectrum
similar to the thermodynamic result can be obtained by inserting a
finite energy resolution, a process mathematically parallel to the
presence of an inelastic scattering rate.  We demonstrate (Section
\ref{ssec-rnd-finite}) some misinterpretations which may arise, we
sketch the scaling procedures, explored in more detail
elsewhere,\cite{janssen:98,mirlin:00} which provide reliable phase
information, and we examine in particular the influence of a finite
energy resolution on the scaling behavior.  Second, we treat the
infinite system within the coherent potential approximation (CPA)
(Section \ref{ssec-cpa}).  This method is known to accurately
reproduce the disorder-averaged DOS while not capturing the
multiple-scattering processes responsible for Anderson localization.
We find (Section \ref{ssec-rnd-infinite}) that the CPA GADOS declines
with increasing disorder
and suggest that this is due to small wavefunction amplitudes at sites of extreme potential.  
In summary, the dependence of
$\rho_g(\omega)$ on disorder strength alone, in the absence of a
clearly defined zero, is not a reliable measure of localization, and
furthermore when the energy resolution is finite, scaling will be
affected.

\section{Calculations}
\label{sec-calc}

\subsection{Model}
\label{ssec-model}

In this work, we will use a site-disordered model, in which the
site energies $\epsilon_i$ take random values, while the hopping
matrix element $t$ is fixed.  For the simplest case of
spinless fermions, the Hamiltonian takes the form
\begin{equation}
H = \sum_{i}|i\rangle \epsilon_i \langle i |
- t \sum_{\langle i,j \rangle }|i\rangle \langle j |
\label{hamiltonian}
\end{equation}
where $\langle i,j\rangle$ indicates nearest neighbour sites on
the lattice, and $|i\rangle$ is a ket corresponding to the Wannier
orbital attached to the $i$th site.  
In this work,
we will present results for box-distributed disorder, where
$\epsilon_i$ are uniformly distributed between $-W/2$ and $+W/2$,
with probability distribution $P(\epsilon) = W^{-1} \Theta(W/2 -
|\epsilon|)$ where $\Theta$ is a step function.  
We will consider two cases: the $d=1$ chain where all
states are known to be localized for any nonzero
$W$,\cite{gogolin:82} and the $d=3$ cubic lattice for which 
all states are localized when $W=16.5$.\cite{mackinnon:81,grussbach:95}

We take two approaches to finding the electronic structure of the
disordered lattice--exact numerical calculation and the coherent potential 
approximation--both of which are standard.

\subsection{Exact numerical calculation (ENC)}
\label{ssec-enc}

The Hamiltonian given in Eq.~(\ref{hamiltonian}) describes noninteracting
Fermions, so that exact single-particle eigenstates may be found by
numerically diagonalizing, using the LAPACK libraries, the matrix
$H_{ij} = \epsilon_i\delta_{i,j} -t\delta_{\langle i,j\rangle}$, where
$\delta_{\langle i,j\rangle}$ is one for nearest neighbor sites and
zero otherwise.  For a system with $N$ sites, we obtain eigenvalues
$E_\alpha$ ($\alpha = 1,\ldots N$) and corresponding eigenvectors
$\Psi_\alpha(\rr_i)$.  We consider systems of linear size $L$
(measured in units of the lattice constant) and dimension $d=1$ or 3,
having $N=L^d$ lattice sites in total.

To calculate the GADOS using the numerical solution, the LDOS is
obtained from the resulting eigenvectors:
\begin{equation}
\rho(\rr_i,\omega) = \sum_\alpha |\Psi_\alpha(\rr_i)|^2
\delta(\omega-E_\alpha).
\label{ldos}
\end{equation}
In numerical calculations, where one has a discrete spectrum, one
generally approximates the delta-function by bins of width $\gamma$:
\begin{eqnarray}
\delta(\omega-E_\alpha) \approx
\gamma^{-1}\Theta(\frac{\gamma}{2}-|E_\alpha-\omega|).
\label{ldosgamma}
\end{eqnarray}
The LDOS approaches the thermodynamic result when $\gamma >
\Delta_N(\omega)$ where $\Delta_N(\omega) = [\rho(\omega) L^d]^{-1} \sim
4dt/N$ is the energy level spacing due to the finite size of the system.  
This binning procedure is essentially equivalent to approximating
\begin{eqnarray}
\rho(\rr_i,\omega) \approx -\pi^{-1} \mathrm{Im }
G(\rr_i,\rr_i,\omega+i\gamma),
\label{imG}
\end{eqnarray}
which is significant because it provides a point of reference between
our results and results for many-body calculations of the
single-particle Green's function.

\begin{figure}
\includegraphics[viewport= 0 00 530 410, width=\columnwidth, clip] {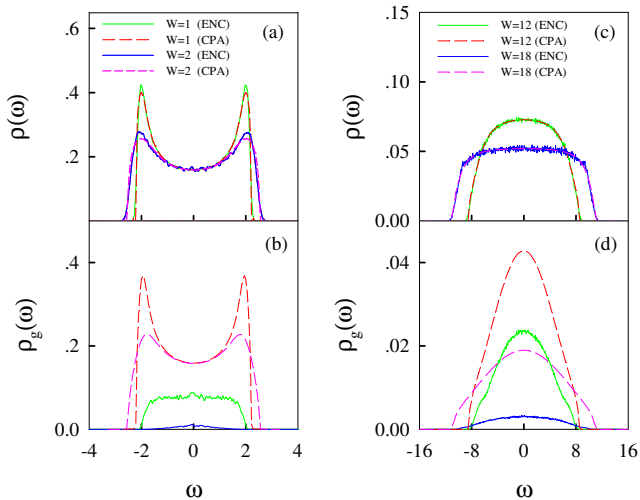}
\caption{(Color online) (a) $d=1$ DOS; (b) $d=1$ GADOS; (c) $d=3$ DOS;
and (d) $d=3$ GADOS.  In each case results from the ENC and from the
CPA are shown for two values of disorder strength.  The ENC results
for $d=1$ are for $N=2000$ averaged over 400 samples, while those for
$d=3$ are for $L=15$ averaged over 200 samples.  All ENC results use
$\gamma=0.05$.}
\label{fig_1}
\end{figure}
The DOS is the arithmetic average of this: $\rho(\omega) =
N^{-1}\sum_{i} \rho(\rr_i,\omega)$, while the GADOS is as defined in
Eq.~(\ref{gados}).  Representative examples of our results for the DOS
and the GADOS in both one and three dimensions are shown in
Fig.~\ref{fig_1}.

We present results both for $\gamma=0$ and $\gamma \ne 0$.  For
$\gamma \ne 0$, we calculate the LDOS using Eq.~(\ref{ldosgamma}).
For $\gamma=0$, we calculate the LDOS using Eq.~(\ref{ldos}) (a
procedure which is only possible for noninteracting systems) in which
case $\rho_g$ is only nonzero for energies corresponding to an energy
eigenvalue.  This result is subsequently smoothed using a binning
procedure.  Note the key distinction that when $\gamma=0$ the binning
is done after the GADOS is calculated.

An important technical issue which must be discussed is disorder
averaging.  In order to improve the signal-to-noise ratio, we
typically present results which have been averaged over several
hundred impurity configurations.  For $\rho(\omega)$ and for the IPR,
this is a meaningful process.  For $\rho_g(\omega)$ it is less clear
whether one should take the arithmetic average or the geometric
average, as was done in Ref.~[\onlinecite{schubert:05}], of different
samples.  We have studied both averaging procedures and have
determined that they yield similar results, except for small values of
$\gamma \lesssim \Delta_N$, for which the geometric averaging becomes
unreliable: If any particular disorder configuration fails to have an
eigenvalue in a particular bin, then $\rho_g(\omega)$ vanishes for
that bin.  For this reason, we find that geometrically averaging
disorder configurations for $\rho_g(\omega)$ exaggerates the already
high sensitivity of the GADOS to single sites with high potential, and
we have adopted the arithmetic average for calculating the disorder
average of $\rho_g(\omega)$.

For purposes of comparison, it is also useful to calculate the
IPR,\cite{bell:70,thouless:74} which is known to be a reliable and sensitive
measure of localization.  For normalized eigenstates, the IPR for
eigenstate $\alpha$ is
\begin{equation}
A_\alpha = \sum_i |\Psi_\alpha(\rr_i)|^4
\label{ipr}
\end{equation}
and it is common to average the IPR over energy bins of width $\gamma$
according to
\begin{equation}
A(\omega) = \frac{
\sum_\alpha \Theta(\frac{\gamma}{2} - |E_\alpha - \omega|) A_\alpha}
{\sum_\alpha \Theta(\frac{\gamma}{2} - |E_\alpha - \omega|)}.
\label{iprgamma}
\end{equation}
It is the scaling of these quantities with system size which
determines whether states are localized: For a system with $N$ sites,
both $A_\alpha$ and $A(\omega)$ scale as $N^{-1}$ for extended states,
but saturate at a finite value when the localization length $\xi$ is
less than the linear dimension $L$ of the system.  In practice when
systems with $L>\xi$ are calculationally inaccessible, localization
can be inferred when $A(\omega)$ vs. $N^{-1}$ extrapolates to a
nonzero value.

\subsection{Coherent potential approximation (CPA)}
\label{ssec-cpa}

The CPA is the second approach we take to determine
the electronic structure of our model.
This is an effective medium theory, modelling a randomly distributed impurity
potential $\epsilon_i$ with a self-energy $\Sigma(\omega)$.  
The self-consistent equation for the self-energy in the CPA
is\cite{schwartz:71}
\begin{equation}
0= \int d\epsilon\, P(\epsilon) \frac{\epsilon - \Sigma(\omega)}
{1-(\epsilon - \Sigma(\omega))G_\mathit{loc}(\omega)}
\label{cpa}
\end{equation}
where the on-site potential is distributed randomly according to the
distribution $P(\epsilon)$, for which we use the box-distribution as
described in Section \ref{ssec-model}.  The disorder-averaged local
Green's function is
\begin{equation}
G_\mathit{loc}(\omega) = \frac{1}{N}\sum_\kk \frac{1}{\omega - t(\kk) - \Sigma(\omega)}
\label{gtilde}
\end{equation}
where, for a tight-binding lattice
\begin{eqnarray}
t(\kk) = \left \{ \begin{array}{cl}
-2t\cos k & d=1 \nonumber \\
-2t(\cos k_x + \cos k_y + \cos k_z) & d=3
\end{array} \right .
\end{eqnarray}
is the dispersion of the lattice.  Equation (\ref{cpa}) is a statement
that when averaged over the effective disorder potential
$\epsilon-\Sigma$ the T-matrix vanishes.

We construct $\rho_g(\omega)$ from the CPA results using the following simple approach:
We find the LDOS for a site embedded in an effective medium 
characterized by the CPA self-energy $\Sigma(\omega)$, where $\Sigma(\omega)$
is determined from Eq.~(\ref{cpa}).
We call this LDOS $\rho_\epsilon(0,\omega)$ because the site is defined to sit at the origin and the site potential is $\epsilon$.
We then take the geometric average over an ensemble of systems where
$\epsilon$ satisfies the 
same distribution function $P(\epsilon)$ used in the CPA calculation:
$\ln [ \rho_g^\mathrm{CPA}(\omega) ] = \int d\epsilon P(\epsilon) \ln[ \rho_\epsilon(0,\omega)]$.
Note that if we take the arithmetic average instead 
we simply recovered the CPA DOS, ie.\ $ \int d\epsilon P(\epsilon) \rho_\epsilon(0,\omega) = \rho^\mathrm{CPA}(\omega) 
 \equiv -\mbox{Im }G_\mathit{loc}(\omega)/\pi$.

Specifically, we solve
the equations of motion for the Green's function with the site energy $\epsilon$ 
at the origin,
\begin{eqnarray*}
{}[\omega^+ - \epsilon]G(0,0,\omega)
- t \sum_n \delta_{\langle 0,n\rangle} G(\rr_n,0,\omega) & = & 1\\
{}[\omega^+ - \Sigma(\omega) ]G(\rr_i,\rr_\ell,\omega)
- t \sum_n \delta_{\langle i,n\rangle} G(\rr_n,\rr_\ell,\omega)
& = & \delta_{i,\ell}
\end{eqnarray*}
where the case $i=\ell=0$ is excluded from the second equation and
$\omega^+=\omega+i0$.   
These give 
\begin{equation}
G(0,0,\omega) = [1+G_\mathit{loc}(\omega)
(\Sigma(\omega)-\epsilon)]^{-1} G_\mathit{loc}(\omega)
\end{equation}
with $G_\mathit{loc}(\omega)$ defined in Eqn.~(\ref{gtilde}).  We remark that
if one imposes the self-consistency condition $G_\mathit{loc}(\omega) = \int d\epsilon P(\epsilon) G(0,0,\omega)$ then the procedure we have outlined is equivalent to the CPA.
Noting that $\rho_\epsilon(0,\omega) =
-\textrm{Im } G(0,0,\omega)/\pi$, 
we then find the GADOS via
\begin{equation}
\ln \rho_g^\mathrm{CPA}(\omega) = \int d\epsilon\, P(\epsilon)
\ln \left [ -\frac 1\pi \textrm{Im } G(0,0,\omega) \right ].
\end{equation}
The CPA results for the $\rho(\omega)$ and $\rho_g(\omega)$ are compared
with exact numerical results in Fig.~\ref{fig_1}.  While the CPA reproduces the
exact $\rho(\omega)$ accurately, there is a significant discrepancy between
$\rho_g^\textrm{CPA}(\omega)$ and $\rho_g^{ENC}(\omega)$, presumably because 
the CPA does not capture the quantum interference to which the LDOS and
hence GADOS are sensitive.   Nonetheless, the CPA
captures the general trend that $\rho_g(\omega)$ is a decreasing function of $W$.
We will explore this point in more detail below 
(Section \ref{ssec-rnd-infinite}). 

\section{Results and discussion}
\label{sec-rnd}

\subsection{Finite systems}
\label{ssec-rnd-finite}

We begin with our results on finite size systems in one and three dimensions.
It may be useful to keep in mind three energy scales and three corresponding length scales.
$\Delta_{\xi}$ is a measure of the average level spacing in the LDOS
caused by localization.  Larger values of $\Delta _{\xi}$ correspond
to stronger localization and shorter localization length:
$\Delta_{\xi} \sim 1/\xi^d$.  
Both $\xi$ and $\Delta_{\xi}$
depend on $\omega$.  $\Delta_N$ is a measure of the mean level spacing
in the DOS caused by the finite size of the system.  Larger values of
$\Delta_N$ correspond to smaller system sizes: $\Delta_N \sim 1/N =
1/L^d$.  Finally, $\gamma$ is the size of the energy bin width used in
numerical calculations of the LDOS [Eq.~(\ref{ldosgamma})] and the IPR
[Eq.~(\ref{iprgamma})].  As suggested by Eq.~(\ref{imG}), $\gamma$
plays the same role as an inelastic scattering rate and the
corresponding length scale is therefore an effective inelastic mean
free path, $\ell_{\gamma} 
\sim 1/\gamma^{1/d}$.\cite{caveat}

\begin{figure}
\includegraphics[width=\columnwidth]{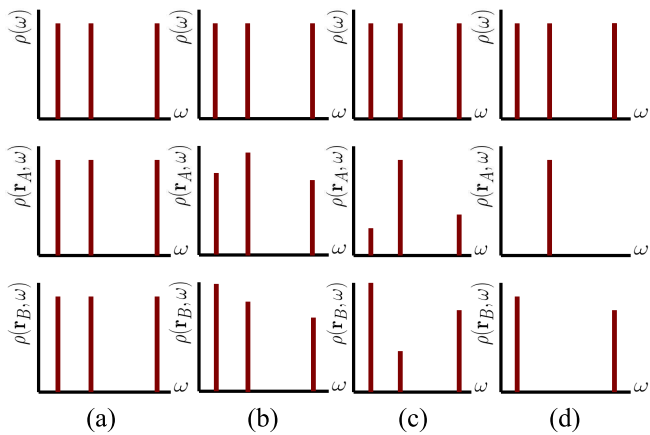}
\caption{(Color online) Sketches of the DOS and LDOS at two
well-separated sites $A$ and $B$ in a sample of fixed size $L$ with 
varying levels of disorder: (a) no disorder, (b) $W<W_c$, (c) $W>W_c$
but $\xi>L$, and (d) $W>W_c$ and $\xi<<L$ and in particular
$\xi<<|{\bf r}_A-{\bf r}_B|$.}
\label{fig_2}
\end{figure}
Fig. \ref{fig_2} presents a useful framework for considering the
effects of finite size on calculations of $\rho_g$.  Four cartoons are
presented corresponding to four values of disorder strength, all in a
finite-size system of length $L$ with $\gamma=0$: (a) no disorder, (b) weak disorder
such that no localization takes place, (c) intermediate disorder such
that localization occurs but with localization length large compared
to the system size, and (d) strong disorder such that the localization
length is much less than the system size.  In each case a sketch is
given within some frequency window of the DOS (for the whole system)
and the LDOS at two well-separated sites $A$ and $B$.  
The lengths of the lines correspond to the weights of the delta-function
peaks in the DOS and LDOS at the discrete eigenenergies of the
system. 
For a system with no disorder, case
(a), the states will be not only extended but also, at low energies,
fairly uniform so the 
weights of the eigenstates
at a given site will be
roughly the same.  When disorder is introduced but no localization
occurs, case (b), the 
weights
at each site are shifted as
sites become more and less energetically favorable.  When localization
begins but $\xi>L$, case (c), the unevenness of the 
weights in the LDOS 
simply
becomes more pronounced.  Only when localization is strong enough that
$\xi<<L$, case (d), do some of the 
weights become exponentially small.

Presenting our results, we begin with the case of a one dimensional
chain, which is useful for several reasons.  First, it is simple:
there is no Anderson transition.  All states are strongly localized
for any nonzero $W$, with $\xi \approx \pi \ell$ where $\ell$ is the
mean free path.  \cite{lee:85,gogolin:82} Furthermore, the limits $L
<\xi$ and $L > \xi$ are both accessible in numerical computations.
Finally, the fact that we can study instances of localization with $W$
less than the bandwidth allows us to 
distinguish two distinct factors 
which cause a decline in the GADOS--exponential tails of localized
eigenstates and 
small wavefunction amplitudes at sites of extreme potential.

\begin{figure}
\includegraphics[viewport= 0 00 500 300, width=\columnwidth, clip ]{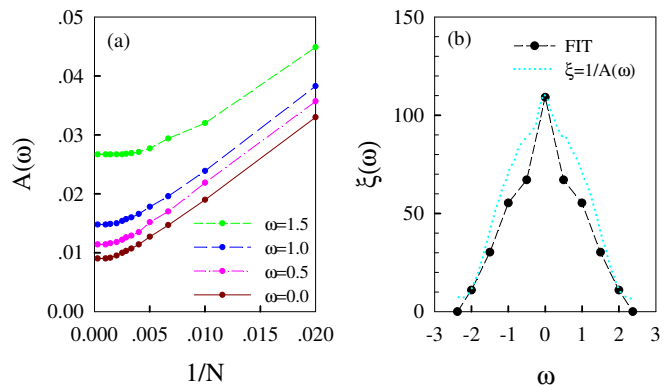}
\caption{(Color online) (a) The inverse participation ratio as a
function of $N^{-1}$ at four different energies for a $d=1$ chain with
$W=1$ and $\gamma = 0.1$.  
(b) The localization length
extracted by a fit to $A(\omega,N)$ (see text), and an alternative definition
$A^{-1}(\omega,N_{max})$ of the localization length 
($N_{max} = 2000$). 
The IPR is averaged over 400 disorder configurations for each
value of $N$. Energies are measured in units of the hopping integral
$t$.}
\label{fig_3}
\end{figure}
Fig.~\ref{fig_3} demonstrates how the IPR shows localization in this
$d=1$ case.  In Fig.~\ref{fig_3} (a), there is a clear crossover from
$A\sim 1/N$ to $A$ roughly constant at $L = \xi$.  The corresponding
energy-dependent localization lengths are shown in Fig.~\ref{fig_3}
(b).
The localization lengths have been extracted by fitting the IPR to a form
$A(\omega,N) = A(\omega,\infty) \coth(N/2\xi)$ which we expect 
for one-dimensional wavefunctions that are decaying exponentially over
a length scale $\xi$ with periodic boundary conditions.  
Because $A(\omega,N)$ saturates for $L\gg\xi$, 
$A(\omega,\infty) \approx
A(\omega,N_{max})$, where  $N_{max}=2000$.
The values of $\xi(\omega)$ determined in this way are quantitatively 
consistent with earlier calculations,\cite{casati:92} and are also 
close to the quantity $A^{-1}(\omega,\infty)$ 
that is frequently taken as a definition of the localization length.
The quantitative differences between the two definitions are not important for
this work.  Rather, the main point of Fig.~\ref{fig_3} is to establish that 
the $d=1$ results presented below are for the strongly-localized regime $L\gg \xi$,
where we expect $\rho_g(\omega)$ to be strongly influenced by the exponential
tails of the wavefunctions.

\begin{figure}
\includegraphics[viewport= 0 00 430 320, width=\columnwidth, clip] {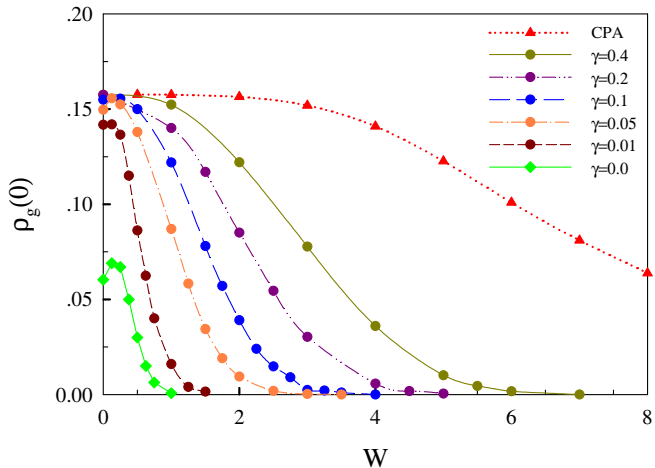}
\caption{(Color online) Dependence of the GADOS at 
$\omega = 0$ on 
$W$ in one dimension, showing ENC results for several bin widths and
CPA results.  
Calculations are in the strongly-localized regime $N=L=2000\gg \xi$ 
and the results are averaged over 400
disorder configurations.  
Energies are measured in units of the hopping integral $t$.}
 \label{fig_4}
\end{figure}

Fig.~\ref{fig_4} shows the variation of the GADOS at 
the band center
as a function of disorder strength for a finite size system.  Results
from the ENC are shown for several values of the energy bin widths
including $\gamma=0$, and the CPA result is also shown.  
Na\"ively, there appears to be an Anderson
transition at nonzero $W$.  If it were to be 
assumed
that the Anderson transition occurs when $\rho_g(\omega) = \eta
\rho(\omega)$ with $\eta \ll 1$, then
$\rho_g(\omega)$ erroneously predicts an Anderson transition at a {\em
nonzero} critical disorder $W_c^\mathit{eff}$, where $W_c^\mathit{eff}$ is strongly
$\gamma$ dependent.  
This happens for two reasons:  First, when $\gamma > \Delta_{\xi}$, the discrete nature of the LDOS is hidden, and moreover when $L<\xi$ even exponentially localized states have significant amplitude at all sites.
The resulting potential for errors points both to the importance of performing a proper scaling analysis and also to the significance of $\gamma$ in the scaling, a feature key to the applicability of this method in many-body calculations.

We 
remark that the $\gamma$ dependence of
$\rho_g(\omega)$ at small values of $W$ in Fig.~\ref{fig_4} is a finite size effect due to
the varying number of discrete states which fall in each energy bin
[see Eq.~(\ref{ldosgamma})].  This issue does not arise when $W$ is
large because in this case the eigenvalues shift significantly from
one disorder configuration to the next and hence the number which fall
in any given energy bin is stabilized by sample averaging.
Equivalently, finite size effects appear for small $W$ because the
mean free path is larger than the system size.

\begin{figure}
\includegraphics[viewport= 0 00 480 320, width=\columnwidth, clip]{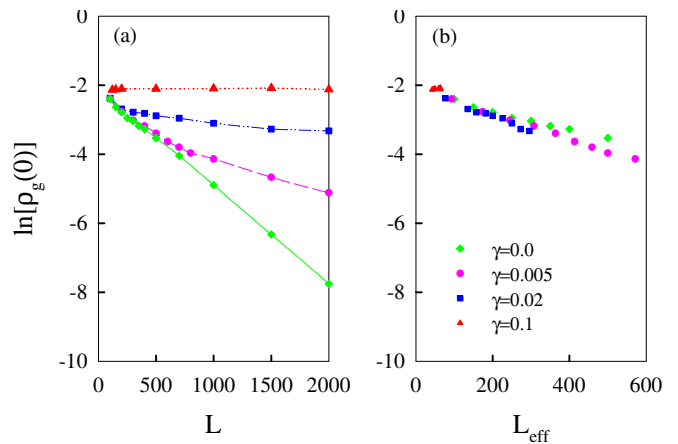}
\caption{(Color online) (a) The scaling with chain length of the ENC GADOS
at the band center 
in one dimension with $W=1$ for several bin widths including
zero.  Results are averaged over 400 disorder configurations.
(b) The same data as in (a), plotted agains $L_\mathit{eff}$.}
\label{fig_5}
\end{figure}

In Fig.~\ref{fig_5}, we examine the scaling behavior of $\rho_g(\omega)$. 
Ref. [\onlinecite{janssen:98}] has done a detailed scaling analysis 
of the critical regime $L < \xi$
for the case $\gamma=0$, i.e. where individual wavefunctions are
studied.  Our goal is to 
study the influence of a finite energy resolution, a necessary
feature of a number of many-body approaches, on the scaling behavior.
Figure~\ref{fig_5}(a) shows the scaling of $\rho_g(0)$ with chain length for $W=1$ and several values of $\gamma$ including zero.
At this disorder strength, $\xi(0) \approx 120$ lattice spacings
 and $\Delta_{\xi}(0) \approx 0.05 t$. For $\gamma=0$ and $L > \xi$, 
as the system size increases the exponentially decaying form of the
localized states manifests itself, and
 $\rho_g(0) \sim \rho(0)\exp(-L/2\xi)$.
  (This form is derived  using periodic 
 boundary conditions).  For $\gamma=0$ and $L<\xi$, the system is in a
critical regime where the decline has a power law rather than
exponential form;\cite{janssen:98} however the difference is difficult to
distinguish over this short scale.
For nonzero $\gamma$, $\rho_g$ scales exponentially for 
$L < \ell_\gamma$
 but saturates at larger values of $L$.  
 We make the ansatz that there is an effective system size
  dominated by the shorter of the two relevant lengths:
 \begin{equation}
 L_\mathit{eff} = ( 1/L + 1/\ell_\gamma)^{-1},
 \label{Leff}
 \end{equation}
 where $\ell_\gamma =  (\rho \gamma)^{1/d}$, such that $\rho_g(0) \sim
 \rho(0) \exp(-L_\mathit{eff}/2\xi)$.  
 In Fig.~\ref{fig_5}(b), a plot of the data from Fig.~\ref{fig_5}(a)
 versus $L_\mathit{eff}$ 
 shows that the ansatz works well.  One immediate consequence
 of Eq.~(\ref{Leff}) is that when $L \gg \ell_\gamma$, $\rho_g(0)$
 saturates at $\sim \rho(0) \exp(-\ell_\gamma/2\xi)$.  Thus, in Fig.~\ref{fig_4}, the apparent critical disorder $W_c^\mathit{eff}$ actually marks
 the crossover from $\xi > \ell_\gamma$ to $\xi < \ell_\gamma$ rather than a
 localization transition.
We conclude that the scaling of $\rho_g(\omega)$ with
$N$ can be used to place an upper bound on the disorder strength at
which localization occurs: If an exponential decay of $\rho_g(0)$ with
increasing $N$ is seen, then localization is occurring.  However, if
$\gamma$ is too large, the effect will be masked.  Essentially this
analysis may lead to false negatives but not to false positives, and
in particular the true result is approached as $\gamma \rightarrow 0$.

We turn our attention now to the $d=3$ cube with $N=L^3$ sites, and $L
\leq 20$.  This case differs from $d=1$ in two important ways.  First,
there is known to be an Anderson transition which, for $\omega=0$
occurs at $W_c=16.5t$.  Second, unless $W$ is very large, we are
generally restricted to $L\ll\xi$.

\begin{figure}
\includegraphics[viewport= 0 00 430 370, width=\columnwidth, clip] {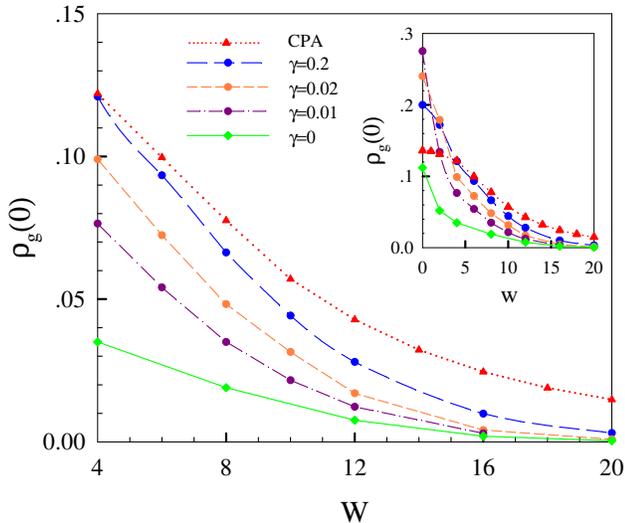}
\caption{(Color online) Dependence of the GADOS at the 
band center on
$W$ (in units of $t$) in three dimensions showing ENC results for
several bin widths and CPA results.  Inset includes results for weak
disorder where there is a strong $\gamma$ dependence.  $L=15$, $N=L^3$
and the results are averaged over 200 disorder configurations.}
\label{fig_6}
\end{figure}
Consider first the variation of the GADOS at the band center as a
function of disorder strength, Fig.~\ref{fig_6}.  
As in the $d=1$ case, the ENC results are $\gamma$ dependent, increasing with increasing $\gamma$.  
Unlike in the $d=1$ case, they are similar to the CPA result, a point discussed further in Section \ref{ssec-rnd-infinite}.
The weak disorder results are included in the inset and show strong $\gamma$ dependence for the same reasons discussed in the $d=1$ case.

\begin{figure}
\includegraphics[viewport= 0 00 500 320, width=\columnwidth, clip] {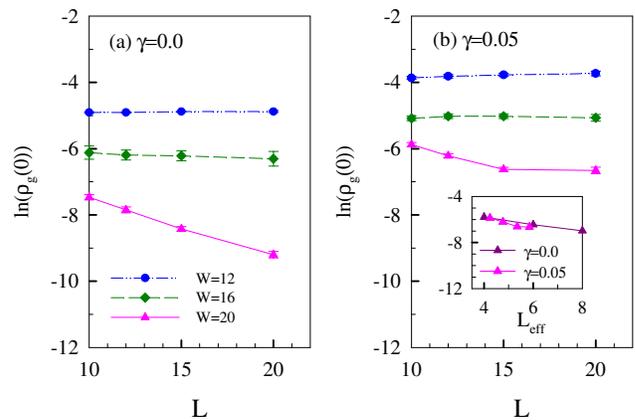}
\caption{(Color online) The scaling with system size of the ENC GADOS
at the band center
in three dimensions, showing disorder strengths
above and below the localization transition, $W_c=16.5$.  (a)
$\gamma=0$ and (b) $\gamma=0.05$.  
The inset shows the scaling with $L_\mathit{eff}$ for $W=20$.
Results are averaged over 200
disorder configurations.  Energies are measured in units of the
hopping integral $t$.}
\label{fig_7}
\end{figure}
Again, 
the correct approach is a scaling analysis.
Fig.~\ref{fig_7} (a) shows the logarithm of the GADOS versus the
system size $L$ ($N=L^3$) using $\gamma=0$, and Fig.~\ref{fig_7} (b)
shows the same quantity for a nonzero energy bin width.  From these
plots it is easy to distinguish localized from delocalized behavior,
at least outside of the critical regime $L < \xi$.
In the delocalized phase, there is no significant variation with
system size.
This is consistent with the log-normal distribution of the LDOS
predicted by Mirlin, \cite{mirlin:94,mirlin:00} and furthermore
the decline in the value of $\rho_g(0)$ with increasing $W$ is
consistent with a decrease in the position of the maximum of the
distribution as the transition is approached, also predicted by
Mirlin.  \cite{mirlin:94,mirlin:00} In the localized phase, the
exponential decay as well as the saturation for $\gamma \ne 0$ are
similar to what is seen in one dimension.  
The larger error bars on
the $\gamma=0$, $W=16$ results are related to the proximity to the
transition: The mobility edges are very close to $\omega=0$ and hence
the energy window used in generating $\rho_g^{\gamma=0}(0)$ must be
especially narrow to avoid including localized states outside the
mobility edge.  The reduction in the number of states included leads
to noisier results.

In one dimension, we showed 
that the strong dependence of $\rho_g(\omega)$ on $\gamma$ can be understood
in terms of an effective length $L_\mathit{eff}$.
We find this to be true for $d=3$ as well. The inset to Fig.~\ref{fig_7}(b)
shows that for the strongly-localized phase ($W=20$), 
the data for $\gamma=0$ and $\gamma=0.05$ collapse 
onto a single curve when plotted against $L_\mathit{eff}$ [Eq.~(\ref{Leff})].
We remark that for $\gamma=0.05$, $\ell_\gamma \approx 8$ lattice constants.
Since $L_\mathit{eff} < \ell_\gamma$, this means that even for a large range
of $L$, the effective range sampled by scaling is quite small.  In particular, if this
re-scaling of lengths applies to the critical regime, this implies that the 
determination of the mobility edges by scaling analysis may be severly limited
in, for example, many-body calculations which are performed at complex frequencies
$\omega + i\gamma$.

\begin{figure}
\includegraphics[viewport= 0 00 400 370, width=\columnwidth, clip] {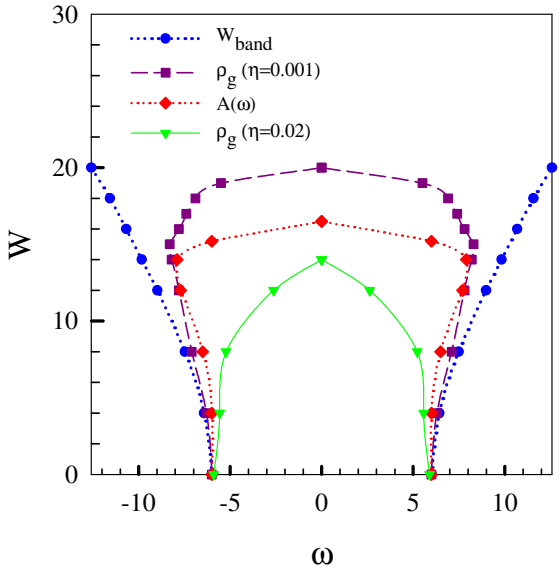}
\caption{(Color online) The bandwidth $W_{band}$ and the mobility
edges $\omega_{c\pm}$ in three dimensions, with the latter estimated
using the scaling of the IPR and $\rho_g(\omega)<\eta \rho(\omega)$
for two values of $\eta$. Further details provided in the text.}
\label{fig_8}
\end{figure}
Finally, Fig.~\ref{fig_8} compares three phase diagrams for $d=3$.
First, we have shown published\cite{bulka:85} results for the mobility edges
based on the scaling of the IPR as a standard of comparison.  
The second
method used is to call states at a given energy localized when the
curve of $\rho_g(\omega)$ versus $W$, like that in Fig.~\ref{fig_6},
drops below some cutoff $\rho_g(\omega) = \eta \rho(\omega)$.  (System
size $L=15$ was used, averaged over 100 to 200 samples, with
$\gamma=0.05$.)  We show curves using two values of $\eta$: one which
results in a localization transition at smaller $W$ than the IPR
transition, and another which results in a localization transition at
higher disorder strength than the IPR transition.  The scaling of
$\rho_g(\omega)$ with system size could also be used to produce a
phase diagram, and the transition would always lie on the high $W$
side of the IPR line, with the two coinciding in the limit $\gamma
\rightarrow 0$.  However, the construction of such a diagram is
similarly computationally intensive to the equivalent IPR calculation
and in an interacting system offers no obvious advantages.  

\subsection{Infinite systems}
\label{ssec-rnd-infinite}

We now turn our attention to infinite systems.
Recent DMFT work
\cite{
dobrosavljevic:97,
aguiar:03,
dobrosavljevic:03,
byczuk:05}
includes plots similar to those in Fig.~\ref{fig_6}, showing the GADOS declining as a function of $W$.
Here, we point out that factors other than exponentially decaying tails of localized
states may lead to a decline (although not a zero) in the GADOS.

The GADOS is very sensitive to small values of the LDOS on individual
sites.  Small values may occur due to localization, but they may also
occur due to the reduction in amplitude of extended wavefunctions at
sites where the magnitude of the potential is large.
The CPA is a convenient tool for examining this issue because it is
known to include the latter effect but not the former.
We show the CPA results for $\rho_g(0)$ as a function of $W$ for $d=1$
in Fig.~\ref{fig_4} and for $d=3$ in Fig.~\ref{fig_6}.  In both cases
$\rho_g^{CPA}(0)$ declines at large $W$.  
In the $d=1$ case, the fact that $\rho_g^\mathrm{CPA}(0)$ is roughly constant
over the range $0<W < 4$ suggests that the suppression of $\rho_g$ in the
ENC calculations is due primarily to the exponentially decaying tails of the localized
eigenstates. 
For $d=3$ the decline in $\rho_g^\mathrm{CPA}(0)$
is especially steep and fairly close to the ENC results,
suggesting that in contrast to the $d=1$ case the $W$ dependence of
$\rho_g(\omega)$ is dominated by 
lattice sites with extreme potential, rather than by exponential tails of localized
eigenstates. 
Therefore, even in an infinite system, a decline in $\rho_g(\omega)$ with
increasing $W$ does not necessarily imply the approach to a localization transition.
This does not appear to change the central conclusions of Refs. \onlinecite{
dobrosavljevic:97,
aguiar:03,
dobrosavljevic:03,
byczuk:05}, which are based on zeros in the GADOS.

\section{Conclusions}
\label{sec-concl}

In summary, for finite-size systems we have demonstrated some of the
false conclusions which may be drawn if a scaling analysis is not used
and shown how the use of a finite energy resolution will influence the
scaling.  Moreover, in infinite systems we have demonstrated that the
GADOS can decline as a function of disorder strength without
localization occurring.

\section*{Acknowledgments}
We wish to thank Ilya Vekhter for helpful conversations.  We would
also like to acknowledge support by NSERC and by Trent University.


\begin{thebibliography}{32}
\bibitem[{cav()}]{caveat}
\bibinfo{note}{{Strictly speaking, it is the dephasing rate which determines
  the length cutoff, as discussed in Ref.~[\onlinecite{altshuler:79}]. For our
  qualitative discussion, however, the distinction is not important.}}
\end{thebibliography}

\end{document}